\documentstyle[amssymb,epsf,aps,floats]{revtex}

\newtheorem{theorem}{Theorem}
\newtheorem{acknowledgement}[theorem]{Acknowledgement}

\begin{document}
\title{Stochastic island generation and influence on the effective transport in
stochastic magnetic fields}
\author{M. Vlad$^{1,2}$, F.\ Spineanu$^{1,2}$, J. H. Misguich$^{2}$, and R.\ Balescu$%
^{3}$}
\address{$^{1}$National Institute for Laser, Plasma and Radiation Physics, \\
Association Euratom-NASTI (MER),\\
P.O.Box MG-36, Magurele, Bucharest, Romania{\it \ }\smallskip \\
$^{2}$Association Euratom-C.E.A. sur la Fusion, CEA/DSM/DRFC, \\
C.E.A.-Cadarache, F-13108 Saint-Paul-lez-Durance, France\\
$^{3}$Association Euratom-Etat Belge sur la Fusion, \\
Universit\'{e} Libre de Bruxelles, CP 231, Campus Plaine\\
Boulevard du Triomphe, 1050 Bruxelles, Belgium}
\date{\today}
\maketitle

\begin{abstract}
The transport of collisional particles in stochastic magnetic fields is
studied using the decorrelation trajectory method. The nonlinear effect of
stochastic generation of magnetic island by magnetic line trapping is
considered together with particle collisions.\ The running diffusion
coefficient is determined for arbitrary values of the statistical parameters
of the stochastic magnetic field and of the collisional velocity. The effect
of the stochastic magnetic islands is analysed. PACS numbers: 52.35.Ra,
52.25.Fi, 05.40.-a, 02.50.-r
\end{abstract}

\section{\protect\bigskip Introduction}

The problem of test particle diffusion in stochastic magnetic fields was
studied by many authors \cite{RR}-\cite{VSRM} and important progress was
obtained. However, the general solution was not yet found. Particle
trajectories in a magnetized plasma are determined by three stochastic
processes: the magnetic field, the collisional velocity along magnetic lines
and the collisional velocity perpendicular to the magnetic lines. These
components of the stochastic collisional velocities have very different
effects. There are two important difficulties appearing in this triple
stochastic process. One is related to the parallel collisional velocity
which enters as a multiplicative noise in the equations of motion and the
other with the Lagrangian non-linearity which is determined by the space
dependence of the stochastic magnetic field. Each of these two problems has
been recently studied but only considered separately. The complete model for
particle transport in stochastic magnetic fields could not be analyzed until
now.

The latter difficulty can be eliminated if one restricts the study to
stochastic magnetic fields with small amplitudes and/or large perpendicular
correlation lengths for which the magnetic Kubo number (defined below) is
small. If the perpendicular collisional velocity is neglected, this
quasilinear problem has an exact solution that was obtained by several
methods \cite{BWM}. It shows that the parallel collisional motion determines
a subdiffusive transport across the confining magnetic field with the
running diffusion coefficient $D(t)$ decaying to zero as $D(t)\sim t^{-1/2}.$%
\ It was shown \cite{VSRM} that this subdiffusive transport is due to
collision induced trajectory trapping along the magnetic lines. The parallel
collisional velocity forces the particles to return in the already visited
positions along the magnetic lines and consequently generates long time
Lagrangian correlation of the stochastic magnetic field. If the
perpendicular collisional velocity is taken into account, the transport is
diffusive and the diffusion coefficient was evaluated semi-qualitatively by
several methods \cite{RR}-\cite{Mcomm}.

On the other hand, the Lagrangian non-linearity determined by the
space-dependence of the stochastic magnetic field leads, at large magnetic
Kubo numbers, to magnetic line trapping and generation of stochastic
magnetic islands. This process is mathematically identical with the
trajectory trapping in the $E\times B$ drift motion in an electrostatic
turbulence. The latter was recently studied by means of a new statistical
approach, the decorrelation trajectory method \cite{V98}, \cite{V00}.

The aim of this paper is to study the general problem of collisional
particle diffusion in stochastic magnetic fields in the guiding center
approximation. More specifically, we determine the effect of self-consistent
generation of stochastic magnetic islands on the effective transport. The
running diffusion coefficient is determined for arbitrary parameters of the
stochastic magnetic field and of particle collisions. The decorrelation
trajectory method is used for studying this rather complicated triple
stochastic process.

The paper is organized as follows. The physical model is described in
Section 2. We derive in Section 3 the Lagrangian velocity correlation and
the running diffusion coefficient for arbitrary values of the four specific
parameters and for given Eulerian correlation of the potential. The physical
significance of this general result is then analyzed: the subdiffusive
transport in Section 4, the effect of collisional cross-field diffusion in
Section 5 and the effect of a time variation of the stochastic magnetic
field in Section 6. The conclusions are summarized in Section 7.

\section{The system of equations}

The particle guiding center motion is studied in a magnetic field with a
stochastic component. The magnetic field is taken to be a sum of a large
constant field ${\bf B}_{0}{\bf =}B_{0}{\bf e}_{z}$ and a small fluctuating
field perpendicular to ${\bf B}_{0},$ and depending on the perpendicular
coordinates ${\bf x}\equiv (x,y)$ and on the parallel coordinate $z$ 
\begin{equation}
{\bf B}=B_{0}\left( {\bf e}_{z}+{\bf b}({\bf x},z,t)\right)  \label{b}
\end{equation}
(Here the perpendicular and the parallel directions are defined in relation
to the direction of ${\bf B}_{0}).$ This is the usual slab model of the
confining configuration in a tokamak plasma. Since the reduced magnetic
field is divergence-free, $\nabla \cdot {\bf b}=0,$ its two components can
be determined from a scalar function $\phi ({\bf x},z)$ as 
\begin{equation}
{\bf b}({\bf x},z,t)={\bf \nabla }\times \phi ({\bf x},z,t){\bf e}_{z}.
\label{2}
\end{equation}
The system of equations for guiding center motion is: 
\begin{eqnarray}
\frac{d{\bf x}}{dt} &=&{\bf b}({\bf x},z,t)\eta _{\parallel }(t)+{\bf \eta }%
_{\perp }(t),  \label{1-x} \\
\frac{dz}{dt} &=&\eta _{\parallel }(t).  \label{1-z}
\end{eqnarray}
The three stochastic functions ${\bf b}({\bf x},z,t),$ ${\bf \eta }_{\perp
}(t)$ and $\eta _{\parallel }(t)$ are statistically independent: all cross
correlations are zero. All these stochastic functions are assumed to be
Gaussian, stationary and homogeneous, with zero averages. The
autocorrelation function of the stochastic potential $\phi ({\bf x},z,t)$ is
modeled by: 
\begin{equation}
A({\bf x},z,t)\equiv \left\langle \phi ({\bf 0},0,0)\,\phi ({\bf x}%
,z,t)\right\rangle =\beta ^{2}\lambda _{\perp }^{2}\exp \left( -\frac{z^{2}}{%
2\lambda _{\perp }^{2}}-\frac{x^{2}+y^{2}}{2\lambda _{\parallel }^{2}}%
\right) \exp \left( -\frac{\left| t\right| }{\tau _{c}}\right)  \label{ecp}
\end{equation}
where \ $\beta $ is the mean square value of the reduced magnetic field $%
{\bf b,}$ $\lambda _{\parallel }$ is the correlation length of the potential 
$\phi $ along the main magnetic field ${\bf B}_{0},$ $\lambda _{\perp }$ is
the correlation length in the plane perpendicular to ${\bf B}_{0}$\ and $%
\tau _{c}$ is the correlation time of $\phi $. The autocorrelation tensor of
the reduced magnetic field components $B_{ij}\equiv \left\langle b_{i}({\bf 0%
},0,0)b_{j}({\bf x},z,t)\right\rangle ,$ $i,j=x,y,$\ is determined from $A(%
{\bf x},z)$ as 
\begin{equation}
B_{xx}=-\frac{\partial ^{2}}{\partial y^{2}}A,\quad B_{yy}=-\frac{\partial
^{2}}{\partial x^{2}}A,\quad B_{xy}=\frac{\partial ^{2}}{\partial x\partial y%
}A.  \label{ecb}
\end{equation}
The collisional velocities are modeled by colored noises with the
correlations 
\begin{equation}
\left\langle \eta _{\parallel }(0)\eta _{\parallel }(t)\right\rangle
_{c}=\chi _{\parallel }\nu R(\nu t)  \label{cpar}
\end{equation}
\begin{equation}
\left\langle \eta _{\perp }^{i}(0)\eta _{\perp }^{j}(t)\right\rangle
_{c}=\delta _{ij}\chi _{\perp }\nu R(\nu t)  \label{cperp}
\end{equation}
where $\nu $ is the collision frequency, $\chi _{\parallel }=\lambda
_{mfp}^{2}\nu /2$ is the parallel collisional diffusivity, $\lambda _{mfp}$
is the parallel mean free path, $\chi _{\perp }=\rho _{L}^{2}\nu /2$ \ is
the perpendicular collisional diffusivity and $\rho _{L}$ is the Larmor
radius relative to the reference field. $R(\nu t)$ is a time decreasing
function that is chosen as 
\begin{equation}
R(\nu t)=\exp (-\nu \left| t\right| )  \label{r}
\end{equation}
for the explicit calculations presented in this paper.

We introduce dimensionless quantities with the following units: $\lambda
_{\perp }$ for the perpendicular displacements, $\lambda _{\parallel }$ for
the displacements along the reference magnetic field and $\nu ^{-1}$ for the
time. The perpendicular velocity ${\bf v}={\bf b}\eta _{\parallel }$ is
reduced with $V\equiv \beta \sqrt{\chi _{\parallel }\nu }$, the parallel
velocity $\eta _{\parallel }(t)$ with $\sqrt{\chi _{\parallel }\nu }$ and
the perpendicular collisional velocity ${\bf \eta }_{\perp }(t)$ with $\sqrt{%
\chi _{\perp }\nu }.$ The equations of motion in these dimensionless
variables (denoted by the same symbols as the physical ones) are:

\begin{equation}
\frac{d{\bf x}}{dt}=M\,{\bf b}({\bf x},z,t)\eta _{\parallel }(t)+\overline{%
\chi }_{\perp }^{1/2}{\bf \eta }_{\perp }(t)  \label{1p-x}
\end{equation}
\begin{equation}
\frac{dz}{dt}=\overline{\chi }_{\parallel }^{1/2}\eta _{\parallel }(t).
\label{1p-z}
\end{equation}
Four dimensionless parameters appear naturally in this problem: the
dimensionless perpendicular and respectively parallel diffusivities

\begin{equation}
\overline{\chi }_{\perp }\equiv \frac{\chi _{\perp }}{\lambda _{\perp
}^{2}\nu },\qquad \overline{\chi }_{\parallel }\equiv \frac{\chi _{\parallel
}}{\lambda _{\parallel }^{2}\nu },  \label{chip}
\end{equation}
a dimensionless parameter that contains the effect of the stochastic
magnetic field

\begin{equation}
M=\frac{V}{\lambda _{\perp }\nu }=\frac{\beta \lambda _{\parallel }}{\lambda
_{\perp }}\overline{\chi }_{\parallel }^{1/2},  \label{M}
\end{equation}
and the dimensionless decorrelation time:

\begin{equation}
\overline{\tau }_{c}=\tau _{c}\nu .  \label{tcor}
\end{equation}
We note that the parameter which describes the evolution of the magnetic
lines, the magnetic Kubo number $K_{m}=\beta \lambda _{\parallel }/\lambda
_{\perp },$ appears here as a factor in $M,$ which can be written as $M=K_{m}%
\overline{\chi }_{\parallel }^{1/2}.$

The aim of this calculation is to determine the Lagrangian correlation of
the effective perpendicular velocity 
\begin{equation}
{\bf v(x},z,t{\bf )\equiv b}({\bf x},z,t)\eta _{\parallel }(t)  \label{vperp}
\end{equation}
which leads to the perpendicular effective diffusion coefficient.

\section{Solution by the decorrelation trajectory method}

We use the decorrelation trajectory method following the recent calculations
for the influence of particle collisions on the diffusion in electrostatic
turbulence \cite{V00}. The difference and the supplementary difficulty of
the magnetic problem comes from the structure (\ref{vperp}) of the velocity $%
{\bf v}$ which is the product of two stochastic processes. They are
statistically independent but in the Lagrangian frame they are correlated
through the trajectories due to the space dependence of the magnetic field
fluctuations. The later makes this problem strongly nonlinear. The
trajectories also depend on the collisional velocity ${\bf \eta }_{\perp }$
and thus the velocity ${\bf v}$ is a triple stochastic process in the
Lagrangian frame.

We determine the collisional contributions to the perpendicular
displacement: 
\begin{equation}
{\bf \xi }(t)=\overline{\chi }_{\perp }^{1/2}\int_{0}^{t}{\bf \eta }_{\perp
}(\tau )d\tau  \label{csi}
\end{equation}
and make the change of variable ${\bf x}^{\prime }(t)={\bf x}(t)-{\bf \xi }%
(t)$ in Eq.(\ref{1p-x}), which introduces the collisional displacements in
the argument of the magnetic field fluctuations: 
\begin{equation}
\frac{d{\bf x}^{\prime }}{dt}=M{\bf v}\left[ {\bf x}^{\prime }(t)+{\bf \xi }%
(t),z,t\right] .  \label{11}
\end{equation}
Here ${\bf v}\left[ {\bf x}^{\prime }(t)+{\bf \xi }(t),z,t\right] {\bf =b}%
\left[ {\bf x}^{\prime }(t)+{\bf \xi }(t),z,t\right] \eta _{\parallel }(t)$
is a triple stochastic process.

We calculate first the Eulerian correlation (EC) of ${\bf \ b}\left[ {\bf x}+%
{\bf \xi }(t),z,t\right] $ which is defined as an average over the magnetic
field fluctuations and over the perpendicular collisional velocity. We
calculate the EC of the potential $\widetilde{\phi }({\bf x},z,t)\equiv \phi %
\left[ {\bf x}+{\bf \zeta }(t),z,t\right] $ \ and then derive the EC of the
magnetic field components. 
\[
E\equiv \left\langle \left\langle \phi \left[ {\bf x}_{1}+{\bf \xi }%
(t_{1}),z_{1},t_{1}\right] \,\phi \left[ {\bf x}_{2}+{\bf \xi }%
(t_{2}),z_{2},t_{2}\right] \right\rangle \right\rangle _{c}= 
\]
\begin{equation}
=\left\langle A\left[ {\bf x}_{1}+{\bf \xi }(t_{1})-{\bf x}_{2}-{\bf \xi }%
(t_{2}),\,z_{1}-z_{2},t_{1}-t_{2}\right] \right\rangle _{c}  \label{ec0}
\end{equation}
This average over the perpendicular collisional velocity can be calculated
using the 2-point Gaussian probability density: 
\begin{equation}
E=\int \int {\bf d\xi }_{1}{\bf d\xi }_{2}A\left[ {\bf x}_{1}-{\bf x}_{2}+%
{\bf \xi }_{1}-{\bf \xi }_{2},\,z_{1}-z_{2},t_{1}-t_{2}\right] P_{2}({\bf %
\xi }_{1},t_{1};{\bf \xi }_{2},t_{2})  \label{ec1}
\end{equation}
where $P_{2}({\bf \xi }_{1},t_{1};{\bf \xi }_{2},t_{2})$ is the probability
density for having ${\bf \xi }(t_{1})={\bf \xi }_{1}$ and ${\bf \xi }(t_{2})=%
{\bf \xi }_{2}.$ It is determined as the average over collisions of the
corresponding product of $\delta $-functions: 
\[
P_{2}({\bf \xi }_{1},t_{1};{\bf \xi }_{2},t_{2})=\left\langle \delta \left( 
{\bf \xi }(t_{1})-{\bf \xi }_{1}\right) \delta ({\bf \xi }(t_{2})-{\bf \xi }%
_{2})\right\rangle _{c}. 
\]
This probability can be calculated using the Fourier representation of the $%
\delta $-functions and the cumulant expansion of the resulting exponential.
Since the collisional displacements are Gaussian, only the first two
cumulants appear. One obtains the two-point probability density for the
perpendicular collisional displacements as 
\begin{equation}
P_{2}=\int \int d{\bf q}_{1}d{\bf q}_{2}\exp \left( i{\bf q}_{1}\cdot {\bf %
\xi }_{1}+i{\bf q}_{2}\cdot {\bf \xi }_{2}-\frac{{\bf q}_{1}^{2}\left\langle 
{\bf \xi }^{2}(t_{1})\right\rangle _{c}}{2}-\frac{{\bf q}_{2}^{2}\left%
\langle {\bf \xi }^{2}(t_{2})\right\rangle _{c}}{2}-\left\langle {\bf q}%
_{1}\cdot {\bf \xi }(t_{1})\,{\bf q}_{2}\cdot {\bf \xi }(t_{2)}\right\rangle
_{c}\right)  \label{mperp}
\end{equation}
which, introduced in Eq.(\ref{ec1}), yields 
\begin{equation}
E({\bf x},z,\tau )=\int {\bf d\xi }\,\,A\left( {\bf x}+{\bf \xi },\,z,\tau
\right) P_{\perp }({\bf \xi },\tau )  \label{ectd}
\end{equation}
after using the dependence of the EC on the difference ${\bf \xi =\xi }_{1}-%
{\bf \xi }_{2}$ and performing the integrals over ${\bf q}_{1},$ ${\bf q}%
_{2} $ and ${\bf \xi }_{1}$. Here $P_{\perp }({\bf \xi },\tau )$ is the
one-point probability density for the perpendicular collisional
displacements: 
\begin{equation}
P_{\perp }({\bf \xi },\tau )=\frac{1}{2\pi \left\langle \zeta ^{2}(\tau
)\right\rangle _{c}}\exp \left( -\frac{{\bf \xi }^{2}}{2\left\langle {\bf %
\xi }^{2}(\tau )\right\rangle _{c}}\right) .  \label{pp}
\end{equation}
The MSD for the collisional perpendicular displacements is: 
\begin{equation}
\left\langle \left[ {\bf \xi }(t_{2})-{\bf \xi }(t_{1})\right]
^{2}\right\rangle _{c}=\left\langle {\bf \xi }^{2}(\tau )\right\rangle _{c}=%
\overline{\chi }_{\perp }\Psi (\tau )  \label{msdp}
\end{equation}
where $\tau \equiv \left| t_{2}-t_{1}\right| $ and $\Psi (\tau ),$ the
reduced mean square collisional displacement, is: 
\begin{equation}
\Psi (\tau )=2\int_{0}^{\tau }(\tau -t)R(t)dt=2\left[ \tau +\exp (-\tau )-1%
\right] .  \label{psi}
\end{equation}
Thus the average effect of the perpendicular collisional velocity ${\bf \eta 
}_{\perp }(t)$ consists of the modification of the EC of the magnetic
potential $\phi .$ The EC $A({\bf x},z,\tau )$ is transformed into $E({\bf x}%
,z,\tau )$ [Eq.(\ref{ectd})] gaining a supplementary time-dependence in
addition to the one determined by the finite correlation time of the
stochastic magnetic field.\ As observed in \cite{V00}, $E$ is the solution
of a diffusive equation and the effect of collisions consists in
progressively smoothing out the EC of the magnetic potential and in
eliminating asymptotically the ${\bf x}$ dependence of $E({\bf x},z,\tau ).$
Since the integral over ${\bf x}$ of $E$ is constant, the time dependence
introduced by collisions in Eq.(\ref{ectd}) does not destroy the correlation
but only spreads it out.

We note that the average over the collisional parallel velocity was not
performed at this stage: $z$ is in Eq. (\ref{ectd}) an Eulerian coordinate.

The problem of collisional particle motion in magnetic turbulence (\ref{1p-x}%
), (\ref{1p-z}) is now formally reduced to a doubly stochastic process: 
\begin{equation}
\frac{d{\bf x}}{dt}=M\,\widetilde{{\bf b}}({\bf x},z,t)\eta _{\parallel }(t)
\label{em}
\end{equation}
\begin{equation}
\frac{dz}{dt}=\overline{\chi }_{\parallel }^{1/2}\eta _{\parallel }(t)
\label{emz}
\end{equation}
where $\widetilde{{\bf b}}{\bf (x},z,t)$ is the stochastic magnetic field
generated by the potential $\widetilde{\phi }({\bf x},z,t)$. The effect of
the perpendicular collisional velocity is an additional time-dependence
introduced in $\widetilde{\phi }({\bf x},z,t)$ and the transformation of its
EC from Eq.(\ref{ec0}) into Eq.(\ref{ectd}). The Eulerian correlation of the
components of $\widetilde{{\bf b}}{\bf (x},z,t)$ are determined from the EC
of the potential (\ref{ectd}) by equations similar to (\ref{ecb}).

The Langevin equation (\ref{em}) can be written as $d{\bf x}/dt={\bf v}({\bf %
x},t)$ and thus it is similar to the two-dimensional divergence-free problem
studied in \cite{V98}. The velocity 
\begin{equation}
{\bf v}({\bf x},z,t)\equiv \widetilde{{\bf b}}({\bf x},z,t)\eta _{\parallel
}(t)  \label{vrad}
\end{equation}
has a much more complicated structure being determined by two multiplied
stochastic processes. However, the method developed in \cite{V98} can be
used here: we will follow the same calculation steps as in \cite{V00}.

First, we define a set of subensembles S of the realizations of the
stochastic functions that have given values of the potential $\widetilde{%
\phi },$ of the magnetic field $\widetilde{{\bf b}}$ and of the parallel
velocity $\eta _{\parallel }$ in the point ${\bf x}={\bf 0},$ $z=0$ at time $%
t=0:$ 
\begin{equation}
\widetilde{\phi }({\bf 0},0,0)=\phi ^{0},\quad \widetilde{{\bf b}}({\bf 0}%
,0,0)={\bf b}^{0},\quad \eta _{\parallel }(0)=\eta ^{0}.  \label{s}
\end{equation}
The correlation of the Lagrangian velocity (\ref{vrad}) can be represented
by a sum over the subensembles of the correlations appearing in each
subensemble 
\begin{equation}
L(t)=\int d\phi ^{0}{\bf db}^{0}d\eta ^{0}P({\bf b}^{0},\phi ^{0},\eta
^{0})\left\langle {\bf v}({\bf 0},0,0){\bf v}\left[ {\bf x}(t),z(t),t\right]
\right\rangle _{S}  \label{LS}
\end{equation}
where $P\left( {\bf b}^{0},\phi ^{0},\eta ^{0}\right) =P\left(
b_{1}^{0}\right) P\left( b_{2}^{0}\right) P\left( \phi ^{0}\right) P\left(
\eta ^{0}\right) $ with $P\left( X\right) =\exp \left( -X^{2}/2\right) /%
\sqrt{2\pi }$ \ is the probability of having \ ${\bf b}^{0},\phi ^{0},\eta
^{0}$ at ${\bf x}={\bf 0},$ $z=0$ and $t=0.$ This probability is a product
of individual distributions because the stochastic variables are not
correlated in ${\bf x}={\bf 0},$ $z=0,$ $t=0.$ The point ${\bf x}={\bf 0},$ $%
z=0$ is taken as the initial condition for the trajectories determined from
Eqs. (\ref{em}), (\ref{emz}). Since the initial velocity in the subensemble
S is ${\bf v}({\bf 0},0,0)={\bf b}^{0}\eta ^{0}$ for all trajectories, the
subensemble average in Eq.(\ref{LS}) is $\left\langle {\bf v}({\bf 0},0,0)%
{\bf v}({\bf x}(t),z(t),t)\right\rangle _{S}={\bf b}^{0}\eta
^{0}\left\langle {\bf v}\left[ {\bf x}(t),z(t),t\right] \right\rangle _{S}$
\ and thus the Lagrangian correlation $L(t)$ is determined by the average \
Lagrangian velocities in all subensembles. In order to evaluate these
quantities, we need to calculate the average Eulerian velocity in the
subensemble S, 
\begin{equation}
{\bf V}^{S}({\bf x},t)\equiv \left\langle {\bf v}\left[ {\bf x},z(t),t\right]
\right\rangle _{S}=\left\langle {\bf b}\left[ {\bf x},z(t),t\right] \eta
_{\parallel }(t)\right\rangle _{S}  \label{vms}
\end{equation}
where $\left\langle {\bf ...}\right\rangle _{S}$ is the average over the two
stochastic processes restricted to the realizations in S and $z(t)$ is the
stochastic parallel displacement obtained from Eq.(\ref{emz}) 
\begin{equation}
z(t)=\overline{\chi }_{\parallel }^{1/2}\int_{0}^{t}d\tau \,\eta _{\parallel
}(\tau ).  \label{zed}
\end{equation}
More precisely, $V_{i}^{S}$ is determined by the following conditional
average 
\begin{equation}
V_{i}^{S}=\frac{\left\langle \left\langle b_{i}\left[ {\bf x},z(t),t\right]
\eta _{\parallel }(t)\,\delta \left[ {\bf b}^{0}-\widetilde{{\bf b}}({\bf 0}%
,0,0)\right] \,\delta \left[ \phi ^{0}-\widetilde{\phi }({\bf 0},0,0)\right]
\delta \left[ \eta ^{0}-\eta _{\parallel }(0)\right] \right\rangle
\right\rangle _{c}}{P\left( {\bf b}^{0},\phi ^{0},\eta ^{0}\right) }
\label{vms1}
\end{equation}
Introducing a function $\delta \left[ z-z(t)\right] $ and using the
statistical independence of $\widetilde{{\bf b}}$ and $\eta _{\parallel }$
one can write 
\[
V_{i}^{S}=\int dz\left\langle b_{i}({\bf x},z,t)\,\delta \left[ {\bf b}^{0}-%
\widetilde{{\bf b}}({\bf 0},0,0)\right] \,\delta \left[ \phi ^{0}-\widetilde{%
\phi }({\bf 0},0,0)\right] \right\rangle \frac{1}{P\left( {\bf b}^{0},\phi
^{0}\right) }\times 
\]
\begin{equation}
\left\langle \eta _{\parallel }(t)\,\delta \left[ z-z(t)\right] \delta \left[
\eta ^{0}-\eta _{\parallel }(0)\right] \right\rangle _{c}\frac{1}{P\left(
\eta ^{0}\right) }.  \label{vms2}
\end{equation}
The first average over the stochastic magnetic field represents the
subensemble average of $b_{i}({\bf x},z,t)$ in S and is given by 
\begin{equation}
B_{i}^{S}({\bf x},z,t)\equiv \left\langle b_{i}({\bf x},z,t)\,\right\rangle
_{S}=\left( \frac{\partial }{\partial x_{2}},\;-\frac{\partial }{\partial
x_{1}}\right) \Phi ^{S}({\bf x},z,t)  \label{bms}
\end{equation}
where $\Phi ^{S}({\bf x},z,t)\equiv \left\langle \widetilde{\phi }({\bf x}%
,z,t)\right\rangle _{S}$, the average potential in the subensemble S$,$ is
calculated as in reference \cite{V00} and is the following function of the
parameters $\phi ^{0},$ ${\bf b}^{0}$ of the subensemble and of the EC of
the potential (\ref{ectd}): 
\begin{equation}
\Phi ^{S}({\bf x},z,t)=\phi ^{0}E({\bf x},z,t)+b_{i}^{0}E_{i\phi }({\bf x}%
,z,t)  \label{fim}
\end{equation}
where $E_{i\phi }({\bf x},z,t)=\left\langle b_{i}({\bf 0},0,0)\,\phi ({\bf x}%
,z,t)\,\right\rangle =-\varepsilon _{ij}\frac{\partial }{\partial x_{j}}E(%
{\bf x},z,t).$

The second average in Eq.(\ref{vms2}) over the collisional parallel velocity
can be written using the Fourier representation of the $\delta $-functions
as 
\[
M_{\parallel }\equiv \left\langle \eta _{\parallel }(t)\delta \left[ z-z(t)%
\right] \delta \left[ \eta ^{0}-\eta _{\parallel }(t)\right] \right\rangle
_{c}\frac{1}{P(\eta ^{0})}= 
\]
\begin{equation}
\frac{1}{P(\eta ^{0})}\int \int dkdq\,\exp \left( -ikz-iq\eta ^{0}\right)
\left\langle \eta _{\parallel }(t)\exp \left[ ikz(t)+iq\eta _{\parallel }(0)%
\right] \right\rangle _{c}.  \label{medpar1}
\end{equation}
The average in this equation can be calculated as the derivative with
respect to $a$ of the following average, evaluated in $a=0$ 
\[
\left\langle \exp \left( a\eta _{\parallel }(t)+ikz(t)+iq\eta _{\parallel
}(0)\right) \right\rangle _{c}= 
\]
\begin{equation}
\exp \left[ -\frac{a^{2}}{2}-\frac{k^{2}}{2}\left\langle
z^{2}(t)\right\rangle _{c}-\frac{q^{2}}{2}+iak\left\langle \eta _{\parallel
}(t)z(t)\right\rangle _{c}+iaqR(t)-kq\left\langle \eta _{\parallel
}(0)z(t)\right\rangle _{c}\right]  \label{mo}
\end{equation}
where 
\begin{equation}
\left\langle \eta _{\parallel }(0)z(t)\right\rangle _{c}=\overline{\chi }%
_{\parallel }^{1/2}\int_{0}^{t}d\tau R(\tau )=\overline{\chi }_{\parallel
}^{1/2}{\cal D}(t)  \label{ezo}
\end{equation}
\begin{equation}
\left\langle \eta _{\parallel }(t)z(t)\right\rangle _{c}=\overline{\chi }%
_{\parallel }^{1/2}\int_{0}^{t}d\tau R(t-\tau )=\overline{\chi }_{\parallel
}^{1/2}{\cal D}(t)  \label{ezt}
\end{equation}
\begin{equation}
\left\langle z^{2}(t)\right\rangle _{c}=\int_{0}^{t}\int_{0}^{t}d\tau
_{1}d\tau _{2}R(\left| \tau _{1}-\tau _{2}\right| )=\overline{\chi }%
_{\parallel }\Psi (t).  \label{msd}
\end{equation}
For the correlation $R$ in Eq.(\ref{r}), the reduced parallel running
diffusion coefficient is 
\[
{\cal D}(t)=1-\exp (-t) 
\]
and the reduced mean square parallel displacement is $\Psi (t)$ defined in
Eq.(\ref{psi}), the same as for the perpendicular collisional displacement.

Straightforward calculations lead to the following equation for the parallel
average $M_{\parallel }$ \ref{medpar1}: 
\begin{equation}
M_{\parallel }(z,t)=\left[ \eta ^{0}R(t)-\overline{\chi }_{\parallel }^{1/2}%
{\cal D}(t)\frac{\partial }{\partial z}\right] P^{S}(z,t)  \label{medpar2}
\end{equation}
where $P^{S}(z,t)$ is the probability of having a parallel displacement $z$
at time $t$ taken for the trajectories in the subensemble $S.$ This was
obtained as a Gaussian distribution with an average displacement $%
\left\langle z(t)\right\rangle _{S}$ and a modified dispersion $%
s(t)=\left\langle \left( z(t)-\left\langle z(t)\right\rangle _{S}\right)
^{2}\right\rangle _{S}^{1/2}:$ 
\begin{equation}
P^{S}(z,t)=\frac{1}{\sqrt{2\pi }s(t)}\exp \left[ -\frac{\left[
z-\left\langle z(t)\right\rangle _{S}\right] ^{2}}{2s^{2}(t)}\right] .
\label{ppars}
\end{equation}
The parallel average displacement is the integral of the parallel average
velocity in S 
\begin{equation}
\left\langle \eta _{\parallel }(t)\right\rangle _{S}=\eta ^{0}R(t)
\label{ems}
\end{equation}
and is obtained as 
\begin{equation}
\left\langle z(t)\right\rangle _{S}=\eta ^{0}\overline{\chi }_{\parallel
}^{1/2}{\cal D}(t).  \label{zms}
\end{equation}
The mean square parallel displacement of the trajectories in S is 
\begin{equation}
s^{2}(t)=\left\langle z^{2}(t)\right\rangle -\overline{\chi }_{\parallel }%
{\cal D}^{2}(t)=\overline{\chi }_{\parallel }\left( \Psi (\tau )-{\cal D}%
^{2}(t)\right) .  \label{msds}
\end{equation}
Thus the dispersion \ of the parallel component of the trajectories in a
subensemble S is always smaller than the dispersion of the whole set of
trajectories $\left\langle z^{2}(t)\right\rangle $.\ It grows slowly (as $%
t^{3})$ at small $t$ and at $t\gg 1$ it reaches $\left\langle
z^{2}(t)\right\rangle .$ The parallel running diffusion coefficient in the
subensemble S is ${\cal D}_{\parallel }^{S}(t)=$ $\overline{\chi }%
_{\parallel }{\cal D}(t)\left( 1-R(t)\right) $. It behaves at small time as $%
t^{2}$ and at $t\gg 1$ it is equal to the global diffusion coefficient of
the whole set of trajectories.

The average velocity (\ref{vms1}) in the subensemble $S$ is thus obtained
using Eqs. (\ref{vms2})-(\ref{medpar2}) as 
\begin{equation}
V_{i}^{S}({\bf x},t)=\int dzB_{i}^{S}({\bf x},z,t)M_{\parallel }(z,t).
\label{vmst}
\end{equation}

The next step in the decorrelation trajectory method is to find a
deterministic trajectory ${\bf X}^{S}(t)$ in each subensemble $S$ as the
solution of the equation 
\begin{equation}
\frac{d{\bf X}^{S}}{dt}=M{\bf V}^{S}({\bf X}^{S},t)  \label{dectr1}
\end{equation}
with ${\bf X}^{S}(0)={\bf 0}.$ Using Eqs.(\ref{vmst}) and (\ref{bms}) \ one
can show that this is a Hamiltonian system of equations which can be written
as: 
\begin{eqnarray}
\frac{dX^{S}}{dt} &=&-M\frac{\partial H^{S}(X^{S},Y^{S},t)}{\partial Y^{S}}
\label{dectr2} \\
\frac{dY^{S}}{dt} &=&M\frac{\partial H^{S}(X^{S},Y^{S},t)}{\partial X^{S}} 
\nonumber
\end{eqnarray}
with the Hamiltonian 
\begin{equation}
H^{S}({\bf X}^{S},t)=\int dz\Phi ^{S}({\bf X}^{S},z,t)M_{\parallel }(z,t).
\label{ham}
\end{equation}
This Hamiltonian represents the average potential in the subensemble S.\ Its
explicit expression calculated for the correlations (\ref{ecp}) and (\ref{r}%
) is: 
\begin{equation}
H^{S}(X^{S},Y^{S},t)=b\eta ^{0}\left( p-n_{\perp }Y^{S}\right) n_{\perp
}\exp \left( -\frac{1}{2}n_{\perp }\left[ \left( X^{S}\right) ^{2}+\left(
Y^{S}\right) ^{2}\right] \right) f_{\parallel }  \label{hex}
\end{equation}
where 
\begin{equation}
f_{\parallel }(t)=n_{\parallel }^{1/2}\left( R-\overline{\chi }_{\parallel
}n_{\parallel }{\cal D}^{2}\left( 1-R\right) \right) \exp \left[ -\frac{1}{2}%
\left( \eta ^{0}\right) ^{2}n_{\parallel }\overline{\chi }_{\parallel }{\cal %
D}^{2}\right] ,  \label{fpar}
\end{equation}
\begin{equation}
n_{\perp }(t)\equiv \left[ 1+\overline{\chi }_{\perp }\Psi (t)\right]
^{-1},\quad n_{\parallel }(t)=\left[ 1+s^{2}(t)\right] ^{-1}.  \label{nuri}
\end{equation}
Since the stochastic magnetic field considered here is isotropic, the
Hamiltonian could be simplified by taking the $x$ axis along ${\bf b}^{0}$.
The parameters of the subensemble S are \ in Eq.(\ref{hex}) $b=\left| {\bf b}%
^{0}\right| ,$ $p\equiv \phi ^{0}/b$ and $\eta ^{0}.$ The equations for the
decorrelation trajectories (\ref{dectr2}) obtain from the Hamiltonian (\ref
{hex}) are 
\begin{eqnarray}
\frac{dX^{S}}{dt} &=&Mb\eta ^{0}n_{\perp }^{2}f_{\parallel }\left[
1+pY^{S}-n_{\perp }\left( Y^{S}\right) ^{2}\right] \exp \left( -\frac{1}{2}%
n_{\perp }\left[ \left( X^{S}\right) ^{2}+\left( Y^{S}\right) ^{2}\right]
\right) ,  \label{dectrx} \\
\frac{dY^{S}}{dt} &=&-Mb\eta ^{0}n_{\perp }^{2}f_{\parallel }X^{S}\left(
p-n_{\perp }Y^{S}\right) \exp \left( -\frac{1}{2}n_{\perp }\left[ \left(
X^{S}\right) ^{2}+\left( Y^{S}\right) ^{2}\right] \right) .  \label{dectr3}
\end{eqnarray}

The average Lagrangian velocity is estimated as in \cite{V00} by the average
Eulerian velocity along the decorrelation trajectory 
\begin{equation}
\left\langle {\bf v}\left[ {\bf x}(t),t\right] \right\rangle _{S}\cong {\bf V%
}^{S}\left[ {\bf X}^{S}(t),t\right]  \label{vmls}
\end{equation}
where ${\bf X}^{S}(t)$ is the solution of Eqs.(\ref{dectrx})-(\ref{dectr3}).

We finally obtain using Eq.(\ref{vmls}) and (\ref{LS}) the correlation of
the perpendicular Lagrangian velocity ${\bf v}\left[ {\bf x}(t),z(t),t\right]
\eta _{\parallel }(t)$ for arbitrary values of the four dimensionless
parameters (\ref{chip})-(\ref{tcor}) and for given Eulerian correlations of
the three stochastic processes that combine in the equations of motion (\ref
{1-x})-(\ref{1-z}): 
\[
L(t;M,\overline{\chi }_{\parallel },\overline{\chi }_{\perp },\overline{\tau 
}_{c})=(\nu \lambda _{\perp })^{2}M^{2}\frac{1}{2\pi }\times 
\]
\begin{equation}
\int_{0}^{\infty }dp\int_{0}^{\infty }db\;b^{3}\exp \left( -\frac{b^{2}}{2}%
(p^{2}+1)\right) \int_{-\infty }^{\infty }d\eta ^{0}\eta ^{0}\exp \left( -%
\frac{\eta ^{02}}{2}\right) V_{1}^{S}\left( {\bf X}^{S}(t),t\right) .
\label{LSF}
\end{equation}
The total perpendicular running diffusion coefficient is the sum of two
terms: a direct contribution of the collisional velocity ${\bf \eta }_{\perp
}$ obtained from Eq.(\ref{msdp}) and the contribution of the velocity (\ref
{vperp}): 
\begin{equation}
D(t;M,\overline{\chi }_{\parallel },\overline{\chi }_{\perp },\overline{\tau 
}_{c})=\chi _{\perp }{\cal D}(t)+(\nu \lambda _{\perp }^{2})MD_{int}(t;M,%
\overline{\chi }_{\parallel },\overline{\chi }_{\perp },\overline{\tau }%
_{c}).  \label{dif}
\end{equation}
The latter is the time-integral of the Lagrangian correlation (\ref{LSF})
and can be written as: 
\begin{equation}
D_{int}=\frac{1}{2\pi }\int_{0}^{\infty }dp\int_{0}^{\infty }db\;b^{3}\exp
\left( -\frac{b^{2}}{2}(p^{2}+1)\right) \int_{-\infty }^{\infty }d\eta
^{0}\eta ^{0}\exp \left( -\frac{\eta ^{02}}{2}\right) X^{S}(t)  \label{dint}
\end{equation}
where $X^{S}(t)$ is the component along $x$ axis of the solution of Eq.(\ref
{dectr2}). It depends on the parameters $M,$ $\overline{\chi }_{\parallel },$
$\overline{\chi }_{\perp }$ and $\overline{\tau }_{c}$ as well as on the
shape of the Eulerian correlations. This contribution (\ref{dint}) results
from the nonlinear interaction of the three stochastic processes. These
results (\ref{LSF})-(\ref{dif}) are written as dimensional quantities.

A computer code that calculates the running diffusion coefficient starting
from the analytical expression (\ref{dint}) was developed. It determines the
decorrelation trajectories (\ref{dectr2}) for a large enough number of
subensembles and performs the integrals in Eq.(\ref{dint}).\ The code was
tested and the parameters in the numerical calculation were established
using the analytical results concerning the subdiffusive transport. Namely,
as shown in the next section, the asymptotic expression for the
decorrelation trajectories and for the diffusion coefficient can be
determined for arbitrary $M$ and $\overline{\chi }_{\parallel }$ if $%
\overline{\chi }_{\perp }=0$ and $\overline{\tau }_{c}=\infty .$ This
provides a very good test for the code and permits the optimization of the
choice of the parameters.

The analyses of the collisional particle transport in stochastic magnetic
fields obtained by means of the decorrelation trajectory method results (\ref
{LSF})-(\ref{dint}) is the subject of the next three sections.

\section{Subdiffusive transport}

We first consider a static stochastic magnetic field ($\tau _{c}\rightarrow
\infty )$ and the zero Larmor radius limit corresponding to negligible cross
field collisional diffusion, $\chi _{\perp }=0.$ It is interesting to study
separately this particular case because it leads to a subdiffusive transport
determined, as shown below, by two kinds of trapping processes. Moreover,
the time dependence of the diffusion coefficient obtained for these
particular conditions allows the understanding of the scaling lows of the
diffusion coefficient determined by the presence of a decorrelation
mechanism.

For the limit $\lambda _{\perp }\rightarrow \infty ,$ an exact analytical
solution was determined \cite{BWM}.\ It was shown that particle
perpendicular transport is subdiffusive with the running diffusion
coefficient going asymptotically to zero as $t^{-1/2}.$ This particular case
is used here as a test for the decorrelation trajectory method. We show that
the exact solution is found. Then the non-linear problem corresponding to
finite $\lambda _{\perp }$ is studied.\ We show that the generation of
magnetic islands by magnetic line trapping does not change the asymptotic
behavior of the diffusion coefficient: a similar subdiffusive regime is
obtained with $D(t)\sim t^{-1/2}.$ The nonlinear process of island
generation has a strong effect but it is localized in time: it determines a
transient \ decrease of $D(t).$ This effect is very important because it
leads, as will be shown in the next sections, to complex anomalous regimes
when $\chi _{\perp }\neq 0$ or when $\tau _{c}$ is finite.

In the limit $\lambda _{\perp }\rightarrow \infty $ the Lagrangian
non-linearity determined by the ${\bf x}$-dependence of the stochastic
magnetic field disappears and the problem simplifies considerably. The
equations for the decorrelation trajectories (\ref{dectr3}) reduce to 
\begin{equation}
\frac{dX}{dt}=-b\eta ^{0}f_{\parallel }(t),\quad \frac{dY}{dt}=0  \label{tro}
\end{equation}
where dimensional quantities were used. Thus the average Lagrangian velocity
in S needed for determining the Lagrangian velocity correlation according to
(\ref{LSF}) is $V_{1}^{S}(t)=-b\eta ^{0}f_{\parallel }(t).$ The integrals
over $p,$ $b$ and $\eta ^{0}$ can easily be performed in Eq.(\ref{LSF}) and
one obtains 
\begin{equation}
L_{0}(t;0,\overline{\chi }_{\parallel },0,\infty )=V^{2}\,\frac{1}{\left[ 1+%
\overline{\chi }_{\parallel }\Psi (t)\right] ^{3/2}}\left[ R(t)\left(
1+s^{2}(t)\right) -\overline{\chi }_{\parallel }{\cal D}^{2}(t)\left[ 1-R(t)%
\right] \right]  \label{Lo}
\end{equation}
which after algebraic transformations becomes 
\begin{equation}
L_{0}(t;0,\overline{\chi }_{\parallel },0,\infty )=\,V^{2}\frac{1}{\left[ 1+%
\overline{\chi }_{\parallel }\Psi (t)\right] ^{1/2}}\left[ R(t)-\frac{%
\overline{\chi }_{\parallel }{\cal D}^{2}(t)}{1+\overline{\chi }_{\parallel
}\Psi (t)}\right] .  \label{Lql}
\end{equation}
This is precisely identical with the exact analytical solution determined in 
\cite{BWM} by means of a different method.\ The perpendicular running
diffusion coefficient can be obtained by time-integration of Eq.(\ref{Lql})
as 
\begin{equation}
D_{0}(t;0,\overline{\chi }_{\parallel },0,\infty )=(V^{2}/\nu )\,\frac{{\cal %
D}(t)}{\left[ 1+\overline{\chi }_{\parallel }\Psi (t)\right] ^{1/2}}.
\label{Dql}
\end{equation}
This exact solution obtained for $\lambda _{\perp }\rightarrow \infty $ is
also valid for finite $\lambda _{\perp }$ if $M=\beta \lambda _{\parallel
}/\lambda _{\perp }\overline{\chi }_{\parallel }^{1/2}\ll 1.$ Actually this
is the condition for neglecting the perpendicular displacements and the $%
{\bf x}$-dependence of the magnetic field fluctuations. Consequently, Eqs.(%
\ref{Lql}), (\ref{Dql}) have physical relevance for tokamak plasmas,
although $\lambda _{\perp }$ is of the order of 1 cm and it is smaller than $%
\lambda _{\parallel }$ by at least a factor $10^{3}.$ Due to the small
values of $\beta $ which are usually of the order $10^{-4}$ the parameter $M$
can be small.

The absolute value of $L_{0}(t;M,\overline{\chi }_{\parallel },0,\infty )$
and $D_{0}(t;M,\overline{\chi }_{\parallel },0,\infty )$ are plotted in
Figures 1 and 2. One can see that the Lagrangian correlation has a long
negative tail at large $t;$ its contribution exactly compensates the
positive part appearing at small time such that its time-integral is zero.
More precisely, $D_{0}\sim t^{-1/2}$ for long time. The zero of the
Lagrangian correlation (and the maximum of $D_{0})$ appears at the average
returning time $\tau _{r}.$ It is determined from the equation $L_{0}(\tau
_{r};M,\overline{\chi }_{\parallel },0,\infty )=0$ and it is a decreasing
function of $\overline{\chi }_{\parallel }$ scaling approximately as $%
\overline{\chi }_{\parallel }^{-1/2}.$ It is remarkable to note that in the
limiting case of absence of collisions ($\nu =0),$ Eq.(\ref{Dql}) yields a
finite diffusion coefficient. In this case, $\chi _{\parallel }\nu
=V_{T}^{2}/2$ (where $V_{T}$ is the thermal velocity) and a small time
expansion can be done in Eq.(\ref{Dql}) obtaining the result of Jokipii and
Parker \cite{JP}, $D_{JP}=\beta ^{2}\lambda _{\parallel }V_{T}/\sqrt{2}.$
This is also well known as the Rochester and Rosenbluth collisionless
diffusion coefficient \cite{RR}, in the form $D_{RR}=D_{m}V_{T},$ where $%
D_{m}$ is the diffusion coefficient of the magnetic lines (see also \cite
{Mag1}, \cite{Mcomm}). Thus, the collisions determine a very strong change
of the perpendicular transport, which is diffusive in the absence of
collisions and becomes subdiffusive due to the parallel collisional motion.
A physical interpretation of this subdiffusive behavior is presented in \cite
{VSRM} in terms of a parallel trapping process determined by the collisions
which force the particles to return in the already visited points along the
magnetic lines. Consequently the Lagrangian velocities remain correlated.
Since the parallel velocity changes its direction due to collisions, this
long-time correlation is negative and thus determines the decay of the
running diffusion coefficient $D_{0}(t).$

A similar subdiffusive transport appears in the non-linear case too,
provided that $\chi _{\perp }=0$. In this case, $n_{\perp }(t)=1$ and the
Hamiltonian (\ref{hex}) depends on time only through the factor $%
f_{\parallel }(t).$ It can be written as: 
\begin{equation}
H(X,Y,t)=f_{\perp }(X,Y)\;f_{\parallel }(t),
\end{equation}
and consequently one can make a change of variable from $t$ to $\tau (t)$
defined by 
\begin{equation}
\frac{d\tau }{dt}=f_{\parallel }(t)  \label{etemp}
\end{equation}
and the equations for the decorrelation trajectories become: 
\begin{equation}
\frac{dX}{d\tau }=-M\frac{\partial f_{\perp }(X,Y)}{\partial Y},\quad \frac{%
dY}{d\tau }=M\frac{\partial f_{\perp }(X,Y)}{\partial X}.  \label{esp}
\end{equation}
The function $\tau (t)$ has a maximum and then decays to zero. The solution
of the time-independent Hamiltonian equations (\ref{esp}) is a periodic
function of $\tau $ with ${\bf X}^{S}(\tau )$ lying on the closed paths
determined by $f_{\perp }(X,Y)=f_{\perp }(0,0).$ The size of the paths
depends only on $p:$ \ it is infinite (straight line) at $p=0$ and decays to
zero as $p$ increases. The period is proportional to $(Mb\eta ^{0})^{-1}.$
The decorrelation trajectories are thus obtained as ${\bf X}^{S}(\tau (t))$
where ${\bf X}^{S}(\tau )$ is the solution of (\ref{esp}). This show that
the trajectories wind around the closed paths (for an incomplete turn or for
many turns, depending on $M$ and on the parameters $p,$ $b$ and $\eta ^{0});$
at the time corresponding to the maximum of $\tau (t)$ they all stop and go
back along the same path. Since $\tau (t)\rightarrow 0$ when $t\rightarrow
\infty ,$ the asymptotic value of the decorrelation trajectories is ${\bf X}%
^{S}(\tau (t))\rightarrow {\bf X}^{S}(0)={\bf 0}.$ All decorrelation
trajectories eventually stop at the origin. The equation for the diffusion
coefficient (\ref{dif}) thus gives $D(t)\rightarrow 0$. Using Eqs.(\ref
{etemp}) and (\ref{fpar}) the function $\tau (t)$ is shown to be $\tau
(t)\cong \left( 2\overline{\chi }_{\parallel }t\right) ^{-1/2}$ at large $t$
and with the solution of Eq.(\ref{esp}) at ${\bf X}^{S}\ll 1$ one obtains $%
X(t)\cong Mb\eta ^{0}\left( 2\overline{\chi }_{\parallel }t\right) ^{-1/2}.$
Upon substitution into Eq.(\ref{dif}) the running diffusion coefficient is
obtained asymptotically as 
\begin{equation}
D(t;M,\overline{\chi }_{\parallel },0,\infty )\rightarrow (\nu \lambda
_{\perp }^{2})M^{2}\left( 2\overline{\chi }_{\parallel }t\right) ^{-1/2}.
\end{equation}
This is identical with the asymptotic behavior obtained from the quasilinear
solution (\ref{Dql}). Thus, the stochastic generation of magnetic islands
that appear at finite $\lambda _{\perp }$ does not affect either the
asymptotic time-dependence of the running diffusion coefficient or its
dependence on the parameters.

There is however a significant effect of the nonlinear process of magnetic
island generation but it appears to be localized in time. It can be found by
determining the whole time evolution of the diffusion coefficient (\ref{dint}%
) using the computer code we have developed. The results are presented in
Figures 1 and 2 compared to the solution (\ref{Lql}), (\ref{Dql}) obtained
for $M\ll 1$. One can see that at small and large times the diffusion
coefficient is equal to $D_{0}(t).$ For intermediary times a transient
decrease of $D(t)$ appears. This is determined by the magnetic line trapping
around stochastic magnetic islands, which is effective at times larger than
the flight time over the perpendicular correlation length $\lambda _{\perp
}, $ which in the unit considered here is $\tau _{fl}=1/M.$ \ As seen in
Figures 1 and 2, the running diffusion coefficient has a maximum at $\tau
_{fl}$ and the Lagrangian velocity correlation becomes negative. Then the
diffusion coefficient decreases due to the trapping of the magnetic lines
which wind around stochastic island. This process is represented by the
decorrelation trajectories corresponding to subensembles with large values
of the parameter $p=\phi ^{0}/b$ \ which have performed many rotations
around their paths (of small size) and their contribution cancels by mixing
in the integrals in Eq.(\ref{dif}). Later in the evolution, another change
of the sign of the Lagrangian correlation is observed at $t=\tau _{r},$ the
average return time for the parallel motion. At this moment $\ D(t)$ has a
minimum while $D_{0}(t)$ has a maximum. It is determined by the parallel
motion and more exactly by the collisions which force the particles to
return on the magnetic lines. This is reflected in the decorrelation
trajectories, which all evolve back on their paths in the perpendicular
plane at $t>\tau _{r}$. In the absence of the magnetic line trapping
(quasilinear conditions) this leads to the decay of the running diffusion
coefficient because the perpendicular displacement decreases in time and
thus $D_{0}(t)$ decays at $t>\tau _{r}$. The effect is inverse in the
presence of stochastic magnetic islands. The backward motion produces first
the un-mixing of the contribution of the trajectories that evolve on trapped
magnetic lines.\ As time increases, the contributions of smaller and smaller
magnetic islands are recovered in the Lagrangian velocity correlation. The
effect of magnetic line trapping that produced the decay of $D(t)$ in the
interval $(\tau _{fl},\;\tau _{r})$ is washed out by the backward motion and 
$D(t)$ recovers its value at $t\sim \tau _{fl}.$ \ At this moment $\tau
_{b}, $ the correlation built-up time, $D(t)$\ has a maximum. A positive
bump appears in the Lagrangian velocity correlation due to the trajectories
unwinding around the magnetic islands. Finally, all decorrelation
trajectories are ''in phase'' and approach the origin. This corresponds to
the asymptotic regime in the evolution of the diffusion coefficient $D(t)$
which is the same as for $D_{0}(t).$ Thus, the parallel collisional motion
eliminates asymptotically the nonlinearity determined by the ${\bf x}$%
-dependence of the magnetic field fluctuations.
\nopagebreak
\begin{figure}[tbp]
\centerline{\epsfxsize=7cm\epsfbox{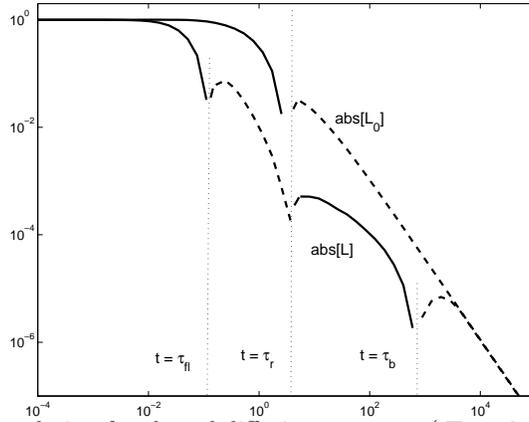}}
\caption{The Lagrangian velocity correlation for the subdiffusive transport ( $%
\overline{\chi }_{\parallel }=0,$ $\overline{\tau }_{c}=\infty ).$ $L_{0}(t)$
corresponds to $M\ll 1$ and is given by Eq.(\ref{Lql}) and $L(t)$ is the
nonlinear result obtained in the presence of magnetic line trapping at large 
$K_{m}$ ($M=10,$ $\overline{\chi }_{\parallel }=0.1).$ The dashed parts of
the two curves represent negative values of the Lagrangian correlations.}
\end{figure}

\begin{figure}[tbp]
\centerline{\epsfxsize=7cm\epsfbox{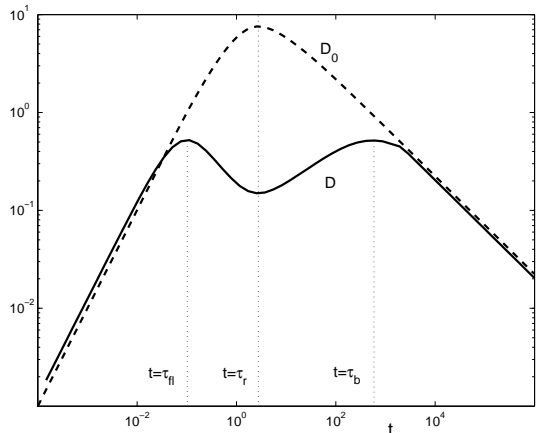}}
\caption{The running diffusion coefficient corresponding to the Lagrangian velocity
correlations in Figure 1: $D_{0}(t)$ is the integral of $L_{0}(t)$ and is
given by Eq.(\ref{Dql}) and $D(t)$ is the integral of $L(t)$ and shows the
effect of the magnetic line trapping. The normalization constant is $\left(
\lambda _{\perp }^{2}\nu \right) M.$}
\end{figure}

The above evolution of the diffusion appears whenever $\tau _{fl}<\tau _{r},$
and since $\tau _{fl}=M^{-1}$ and $\tau _{r}\approx \overline{\chi }%
_{\parallel }^{-1/2},$ the condition is $K_{m}>1$ which corresponds to
magnetic line\ trapping. When $\tau _{fl}>\tau _{r}$ (or $K_{m}\ll 1),$ the
diffusion coefficient is given by Eq.(\ref{Dql}).

We show in the next sections that this rather nontrivial evolution of the
running diffusion coefficient leads to anomalous diffusion regimes when a
decorrelation mechanism is present.

\section{Diffusive transport induced by collisional decorrelation}

We analyze in this section the effect of the cross-field collisional
diffusion ($\overline{\chi }_{\perp }\neq 0)$ starting from the general
solution (\ref{LSF})-(\ref{dint}). The stochastic collisional velocity ${\bf %
\eta }_{\perp }(t)$ in Eq.(\ref{1-x}) moves the particles out of the
magnetic lines and consequently it has a decorrelation effect leading to
diffusive transport. This collisional motion determines a characteristic
time, the perpendicular decorrelation time $\tau _{\perp }.$ It is defined
by the condition that the collisional diffusion covers the perpendicular
correlation length, $2\chi _{\perp }\tau _{\perp }=\lambda _{\perp }^{2},$
and in the units chosen here it is $\tau _{\perp }=(2\overline{\chi }_{\perp
})^{-1}.$ The stochastic magnetic field is considered here to be static ($%
\overline{\tau }_{c}=\infty )$ for a better understanding of the collisional
decorrelation.

As in the previous section, a stochastic magnetic field with small Kubo
number $K_{m}$ that does not generate stochastic magnetic islands is first
considered. We show analytically that the already known results are
reproduced by the decorrelation trajectory method.\ Then the nonlinear case
is analyzed and new anomalous diffusion regimes are found. They are
determined by the non-linear interaction of the magnetic line trapping with
the cross-field collisional diffusion.

In 1979 Kadomtsev and Pogutse \cite{KP} derived semi-qualitatively an
approximation for the cross-field diffusion coefficient. This approximation
is essentially a weak-nonlinearity regime, in which the magnetic field
fluctuations are non-chaotic. It will be shown that this diffusion
coefficient is obtained from the general equations (\ref{LSF})-(\ref{dint})
provided that $\tau _{r}<\tau _{\perp }<\tau _{fl}.$ This condition is
compatible with the relations found in \cite{Mcomm} where a detailed study
of the diffusion regimes in stochastic magnetic fields for fusion plasmas is
presented. In this conditions the ${\bf X}^{S}$-dependence of the average
velocity in Eqs.(\ref{dectrx}), (\ref{dectr3}) can be neglected and the
equations for the decorrelation trajectories are (\ref{tro}) corrected by a
factor $n_{\perp }^{2}(t)$ that multiplies the right hand side terms. This
leads to the following form of the Lagrangian velocity correlation

\begin{equation}
L_{KP}(t)=n_{\bot }^{2}(t)\,L_{0}(t),  \label{eq.36}
\end{equation}
where $L_{0}(t)$ is the subdiffusive Lagrangian velocity autocorrelation
defined in Eq. (\ref{Lql}). Because of the factor $n_{\bot }^{2}(t)$, the
integral of $L_{KP}(t)$ no longer vanishes, and yields a finite diffusion
coefficient, $D_{KP}$. It can be estimated analytically by using a step
approximation of the function $n_{\bot }(t)$

\begin{equation}
n_{\bot }(t)\cong \left\{ 
\begin{array}{c}
1,\;\;\;t<\tau _{\perp } \\ 
0,\;\;\;\;t>\tau _{\perp }
\end{array}
\right.  \label{eq.39}
\end{equation}
It then follows that the diffusion coefficient is approximated as:

\begin{equation}
D_{KP}\cong \int_{0}^{\tau _{\perp }}dt\,L_{0}(t)=-\int_{\tau _{\perp
}}^{\infty }dt\,L_{0}(t)  \label{eq.41}
\end{equation}
because the integral of $L_{0}(t)$ from $t=0$ to infinity is zero. Using the
very simple asymptotic form of $L_{0}(t)$ [obtained from Eq.(\ref{Lql}) for $%
\tau >\tau _{r}$], the integral can be calculated analytically and one
obtains (going to dimensional quantities)

\begin{equation}
D_{KP}\cong \beta ^{2}\,\frac{\lambda _{\Vert }}{\lambda _{\bot }}\sqrt{\chi
_{\Vert }\,\chi _{\bot }}  \label{eq.46}
\end{equation}
which is the well-known Kadomtsev-Pogutse formula.

When the time of flight $\tau _{fl}$ is smaller than the decorrelation time $%
\tau _{\bot },$ the space dependence of the magnetic field fluctuations
cannot be neglected. It leads to stochastic magnetic islands. In the
presence of a perpendicular collisional diffusivity the decorrelation
trajectories obtained from Eq.(\ref{dectr3}) are not more closed curves.
However, trajectory winding can still be observed for some range of the
parameters that define the subensembles. This means that the process of
generation of\ stochastic magnetic island and of magnetic line trapping
still exists. Compared to the decorrelation trajectories obtained with $\chi
_{\perp }=0,$ these trajectories saturate faster and perform a smaller
number of rotations. They still turn back at the maximum of the function $%
\tau (t)$ which shows that the parallel trapping determined by the parallel
collisional motion still exists. But due to the cross field collisional
diffusion, these two trapping processes are only approximate or temporary.
The perpendicular diffusion $\chi _{\bot }$ produces a releasing effect both
for perpendicular and parallel components of particle motion. The asymptotic
values of the decorrelation trajectories are not concentrated in the origin
(as for $\chi _{\bot }=0)$ but spread in the $(X,Y)$ plane. Consequently, a
finite value of the asymptotic diffusion coefficient yields from Eq.(\ref
{dif}).

\begin{figure}[tbp]
\centerline{\epsfxsize=7cm\epsfbox{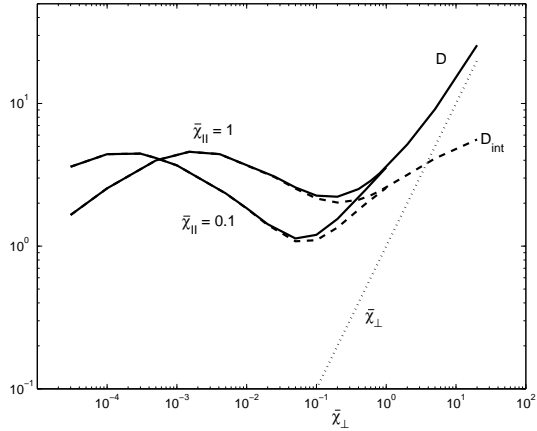}}
\caption{The asymptotic diffusion coefficient as a function of $\overline{\chi }%
_{\perp }.$ The total diffusion coefficient $D$ (continuous lines) is
compared with the direct collisional contribution $\overline{\chi }_{\perp }$
(dotted line) and with the interaction term $D_{int}$ (dashed lines) for two
values of $\overline{\chi }_{\parallel }.$ The normalization constant is $%
\lambda _{\perp }^{2}\nu ,$ $M=10$ and $\overline{\tau }_{c}=\infty .$}
\end{figure}

The asymptotic diffusion coefficient is determined from Eq.(\ref{dif}) using
the numerical code we have developed. Some results are presented in Figure 3
where the asymptotic value of $D(t)$ is represented as a function of $%
\overline{\chi }_{\perp }.$ The two components $D_{int}$ and $\overline{\chi 
}_{\perp }$ are also represented. One can see that at small collisional
diffusion $\overline{\chi }_{\perp }\ll 1$, the non-linear interaction term
largely dominates the collisional term while at large collisional diffusion $%
\overline{\chi }_{\perp }\gtrsim 1,$ the nonlinear term is only a correction
to $\overline{\chi }_{\perp }.$ Thus, the subdiffusive transport appearing
at $\overline{\chi }_{\perp }=0$ is transformed by a small collisional cross
field diffusion into a diffusive transport with a diffusion coefficient that
can be several orders of magnitude larger than $\overline{\chi }_{\perp }.$
The dependence of the diffusion coefficient on $\overline{\chi }_{\perp }$
is rather nontrivial. There is at very small $\overline{\chi }_{\perp }$ an
increase of $D$ up to a maximum which corresponds to $\tau _{\perp }\cong
\tau _{b}.$ Then, at larger $\overline{\chi }_{\perp },$ the nonlinear
interaction of the parallel and perpendicular trapping with the collisional
decorrelation generates a strange transport regime, in which the effective
diffusion coefficient decreases as the collisional diffusion $\overline{\chi 
}_{\perp }$ increases. A minimum of $D$ is obtained when $\overline{\chi }%
_{\perp }$ determines a decorrelation time of the order of the return time
of the parallel motion, $\tau _{\perp }\cong \tau _{r}.$ At larger $%
\overline{\chi }_{\perp }$ (when $\tau _{\perp }<\tau _{r}),$ the nonlinear
contribution $D_{int}$ increases again with the increase of $\overline{\chi }%
_{\perp }$ but this contribution begins to be comparable and eventually
negligible compared to the collisional diffusion coefficient.

We note that the above results obtained with the decorrelation trajectory
method are not similar with the heuristic estimation of the asymptotic
diffusion coefficient of Rechester and Rosenbluth \cite{RR}. This is
possibly due to the fact that the trapping of the magnetic lines, which is
profoundly implied in the above results, is neglected in the estimation \cite
{RR} and also in the more detailed calculations presented in \cite{Mag1}.
This estimation is based on the process of exponential increase of the
average distance between two magnetic lines in a chaotic magnetic field,
represented by the Kolmogorov length. The estimation of this length taking
into account the trapping of the magnetic lines should be necessary in order
to compare the results.

\section{Diffusive transport in time-dependent stochastic magnetic fields}

In a time-dependent stochastic magnetic field with finite $\tau _{c}$ the
configuration of the stochastic field ${\bf b}({\bf x},z,t)$ changes, the
magnetic lines move and consequently the perpendicular velocity of the
particles is decorrelated leading to diffusive transport. We determine here
the diffusion coefficient in such time-dependent fields in the limit of zero
Larmor radius, stating from the general solution (\ref{LSF})-(\ref{dif}).
The effect of time variation of the stochastic magnetic field on the
effective diffusion was previously studied in \cite{td1}-\cite{td4} but only
for weak magnetic turbulence ($K_{m}\ll 1).$ We determine the effect of
stochastic island generation appearing in stochastic magnetic fields at $%
K_{m}>1.$

The decorrelation trajectories obtained from Eqs.(\ref{dectrx}), (\ref
{dectr3}) are in this case (finite $\tau _{c},$ $\overline{\chi }_{\perp
}=0) $ situated on closed paths (except that for $p=0)$. A typical
trajectory rotates on the corresponding path, then it stops and turns back.
Its velocity decays progressively and eventually the trajectory stops
somewhere on its path. This is the modification determined by the time
variation of the magnetic field: all decorrelation trajectories stop at a
time of the order $\overline{\tau }_{c}.$ Consequently, the running
diffusion coefficient saturates. Depending on the relation between the
decorrelation time $\overline{\tau }_{c}$ and the three characteristic times
of this motion, $\tau _{fl},$ $\tau _{r},$ $\tau _{b}$ (see Fig. 2) several
diffusion regimes are obtained. \ In time-dependent magnetic fields, at $%
t<\tau _{c},$ the time evolution of the diffusion coefficient is
approximately the same with that obtained for $\overline{\tau }%
_{c}\rightarrow \infty ,$ and later, at $t>\tau _{c},$ $D(t)$ saturates.
Thus, the asymptotic diffusion coefficient can be evaluated as 
\begin{equation}
\mathrel{\mathop{\lim }\limits_{t\rightarrow \infty }}%
D(t;M,\overline{\chi }_{\parallel },0,\overline{\tau }_{c})\cong D(\overline{%
\tau }_{c};M,\overline{\chi }_{\parallel },0,\infty ).  \label{dtd}
\end{equation}
using the running diffusion coefficient obtained in the static case. Thus it
can be approximating by the value of the running diffusion coefficient for
the subdiffusive case at $t=\overline{\tau }_{c}.$ \ Some results are
presented in Figure 4 where the asymptotic diffusion coefficient obtained
from Eq.(\ref{dint}) \ for finite $\overline{\tau }_{c}$ is compared to the
subdiffusive running diffusion coefficient represented in Figure 2. One can
see that the approximation (\ref{dtd}) is rather good for all values of $\ 
\overline{\tau }_{c}.$

\begin{figure}[tbp]
\centerline{\epsfxsize=7cm\epsfbox{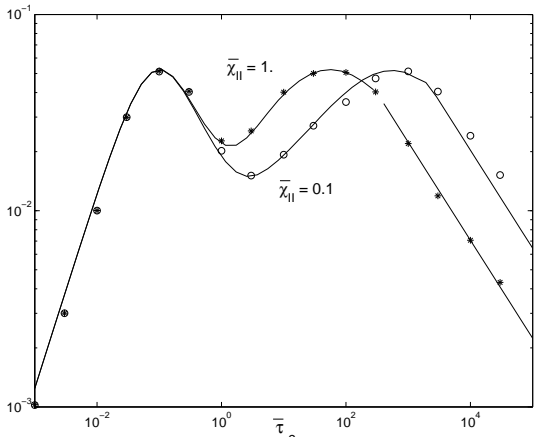}}
\caption{The asymptotic diffusion coefficient as a function of $\overline{\tau }_{c}$
for $\overline{\chi }_{\parallel }=0.1$ (circles) and $\overline{\chi }%
_{\parallel }=1$ (stars). The continuous lines represent the running
diffusion coefficient as a function of $t$ for the subdiffusive transport
corresponding to static magnetic fields ($\overline{\tau }_{c}=\infty ).$
The normalization constant is $\left( \lambda _{\perp }^{2}\nu \right) M^{2}$%
, $M=10,$ $\overline{\chi }_{\perp }=0.$}
\end{figure}

The following diffusion regimes can be observed in Figure 4, in the
nonlinear conditions when stochastic magnetic islands are generated ($\tau
_{fl}<\tau _{r},$ or $K_{m}>1).$ The quasilinear regime at small correlation
times ($\overline{\tau }_{c}<\tau _{fl}$) with $D_{0}\approx M^{2}\overline{%
\tau }_{c}$ is characterized by a fast time-variation which prevents
trajectory trapping. At larger correlation times ($\tau _{fl}<\overline{\tau 
}_{c}<\tau _{r})$ the stochastic magnetic islands can be generated before
the stochastic magnetic field changes and the parallel motion is ballistic.
In these conditions the diffusion regime is similar to that described in 
\cite{V00} for the electrostatic turbulence: the diffusion coefficient
decreases with the increase of $\overline{\tau }_{c}.$ A minimum of the
diffusion coefficient appears at $\overline{\tau }_{c}\cong \tau _{r},$
followed by an anomalous increase determined by the interaction of the
parallel trapping with the magnetic line trapping which generates
correlation of the Lagrangian velocities. At very large correlation times ($%
\overline{\tau }_{c}>\tau _{b})$ the diffusion coefficient decreases as $%
D\approx K_{m}^{2}\overline{\tau }_{c}^{-1/2}\overline{\chi }_{\parallel
}^{1/2}.$ We note that the regimes obtained for $\overline{\tau }_{c}<\tau
_{fl}$ and for $\overline{\tau }_{c}>\tau _{b}$ are similar with those
reported in \cite{td1}, \cite{Mag2}. But instead of the plateau found there
at intermediary $\overline{\tau }_{c},$ we obtain here a more complicated
behavior. This is the effect of stochastic magnetic island generation: it
leads to the decrease of the effective diffusion coefficient with the
increase of $\overline{\tau }_{c}$ when the parallel motion is ballistic
and, on the contrary, to the increase of $D$ with the increase of $\overline{%
\tau }_{c}$ when the parallel motion is diffusive.

\section{Conclusions}

\bigskip We have studied here the transport of collisional particles in
stochastic magnetic fields using the decorrelation trajectory method. We
have derived analytical expressions for the running diffusion coefficient
and for the Lagrangian velocity correlation in terms of a set of
deterministic trajectories. They are defined in subensembles of the
realizations of the stochastic field as solution of differential
(Hamiltonian) equations that depend on the given Eulerian correlation of the
stochastic potential. They are approximations of the subensemble average
trajectories and represent the dynamics of the decorrelation of the
Lagrangian velocity. Since in general the equations for the decorrelation
trajectories cannot be solved analytically, a computer code was developed
for determining the running diffusion coefficient for arbitrary values of
the four parameters of this problem and for given Eulerian correlation of
the potential.

We have shown that this rather complicated triple stochastic process is
characterized by two kinds of trajectory trappings and contains two
decorrelation mechanisms. The latter are produced by the collisional cross
field diffusion $\overline{\chi }_{\perp }$ and by the time variation of the
stochastic magnetic field.

One of the trapping processes concerns the parallel motion and is determined
by collisions which constrain the particles to return in the already visited
places with probability one. This parallel trapping leads to a subdiffusive
transport in the absence of a decorrelation mechanism. This already known
process is recovered by our method. The second kind of trapping concerns the
magnetic lines which at $K_{m}>1$ wind around the extrema of the vector
potential generating self-consistently magnetic islands. The effects of the
magnetic line trapping in the presence of particle collisions is studied for
the first time. We show that in the absence of a decorrelation mechanism,
the stochastic magnetic islands determine a transitory decay of the running
diffusion coefficient $D(t)$ appearing at $t$ in the interval $(\tau
_{fl},\;\tau _{r}),$ i.e. before the parallel trapping is effective. The
simultaneous action of both trapping processes determine a nonlinear built
up of Lagrangian velocity correlation and eventually the parallel motion
washes out the effect of the magnetic line trapping. Consequently, the
asymptotic behavior of the running diffusion coefficient is exactly the same
as in the quasilinear conditions when the stochastic magnetic field does not
generate magnetic islands.

The effect of the two decorrelation mechanisms is afterwards studied. We
show that the effective diffusion coefficient and its dependence on the
parameters results from a competition between the trapping and the
decorrelation processes and more precisely from the temporal ordering of the
characteristic times of these processes. Each one of the two decorrelation
mechanisms leads to the already known diffusion laws when the stochastic
magnetic islands are not present ($K_{m}\ll 1).$ Their presence (at $%
K_{m}>1) $ produces a complicated nonlinear interaction between the three
stochastic processes which determines new scaling laws of the diffusion
coefficient. They appear when the decorrelation time is longer than the
flight time $\tau _{fl}$ but smaller than the correlation built up time $%
\tau _{b}.$ The first condition ensures the magnetic islands generation and
the second prevents the elimination of their trapping effect by the parallel
collisional motion. A particularly interesting regime is obtained for
collisional decorrelation and consists of an effective diffusion coefficient
that decreases when the collisional perpendicular diffusion increases (Fig.
3).

This rather complex dependence of the diffusion coefficients on the plasma
parameters can be used in experiments for controlling the transport. Even
without changing the characteristics of the stochastic magnetic field, the
diffusion coefficient can be strongly influenced by the parameters which
describe particle collisions. A \ minimum of the diffusion coefficient was
obtained for decorrelation times of the order of the average return time for
the parallel motion.

\begin{acknowledgement}
This work has benefited of the NATO Linkage Grant PST.CLG.977397 which is
acknowledged.
\end{acknowledgement}

\end{document}